\documentclass[showpacs,preprintnumbers,amsmath,amssymb]{revtex4}
\allowdisplaybreaks
\usepackage{graphicx}
\input epsf
\usepackage{dcolumn}
\usepackage{hyperref}
\usepackage{showkeys}
\hypersetup{colorlinks=true, citecolor=blue, linkcolor=red}
\newcommand{\beq}{\begin{equation}}
\newcommand{\beqn}{\begin{equation*}}
\newcommand{\enq}{\end{equation}}
\newcommand{\enqn}{\end{equation*}}
\newcommand{\eb}{{\rm e}}
\newcommand{\sech}{{\rm sech}}
\newcommand{\R}{{\mathbb R}}

\renewcommand{\l}{\lambda}

\begin{document}
\allowdisplaybreaks
\title{The KdV in Cosmology: a useful tool or a distraction?}
\author{A. V. Yaparova}%
\email{lisa74@yandex.ru}
\author{A. V. Yurov}%
\email{AIUrov@kantiana.ru}
\author{V. A. Yurov}%
\email{vayt37@gmail.com}
\affiliation{Immanuel Kant Baltic Federal University, Department of Physics and Technology,
 Al.Nevsky St. 14, Kaliningrad 236041, Russia}

\date{\today}
\begin{abstract}
The letter is a response to the recent article by J. Lidsey \cite{L}. We demonstrate that the Schwarzian derivative technique developed therein is but a consequence of linearizabiliy of the original cosmological equations. Furthermore, we show the required linearized equation to be nothing else but a Schr\"odinger equation.

\end{abstract}


\maketitle
\section{Introduction} \label{sec:intro}

It would hardly be a mistake to see a history of an inflationary cosmology as an exceptional succession of brilliant mathematical ideas and concepts. It would be equally correct to say that most of these concepts has originated in the areas of mathematical physics that had little or nothing to do with cosmology. Indeed, if one should try to count every idea that has seen an application in the inflationary cosmology, a corresponding list would stretch from the Riemann zeta function to the supercooling effect \cite{G} to the vacuum energy and phantom fields. Among its items, a special place will be rightfully reserved for the differential equations that appear for various reasons and in various guises -- the Riccati equation, Schr\"oedinger equation, Abel equation... Browsing through them, one is left with an impression that it is perhaps not entirely unnatural to expect some other important differential equations from, say, hydrodynamics or plasma physics, to suddenly pop up somewhere in the studies of an early universe. So, it was not surprising to see an article \cite{L} by James Lidsey claiming that one of the ways to solve the Friedmann equations under the slow-rolling approximation invokes a technique reminiscent of the B\"acklund transformation for the Korteweg-de Vries equation. However, after a due consideration of the matter it is not difficult to realize that the technique in question has little to do with KdV; instead, it is a direct consequence of a linearizability of the equations involved. In fact, we don't need the KdV at all, for the linearized equation ends up being the {\em Schr\"odinger equation}. And this, of course, opens up a range of new possibilities; for example, one can use the Darboux transformation to generate a whole set of new solutions of the Schr\"odinger equation that will describe a slow-rolling inflation with a $\phi$-dependent scalar parameter.

The article is divided in the following manner. In chapter \ref{sec:infl} we discuss an analytic approach to the inflationary cosmology and demonstrate that after the reduction to a first order ODE, the new equation ends up being a generally non-integrable Abel equation of the first kind, therefore prompting a use of a suitable approximation, namely -- a slow-rolling approximation. We than adopt a Lidsey's approach and show how to solve the resulting equations using the Schwarzian differential. However, this very solution can be obtained much faster and in a more intuitively clear fashion once we figure out a proper linearization of the approximate Friedman equations. Such a linearization is derived in the first part of section \ref{sec:Sch}. Then we discuss the further benefits of the new representation. And in the conclusion of the article we give our answer to a hypothetical question: does there indeed exists an unambiguous connection between the Friedman and the KdV equations. In short, the answer is: {\em yes}.

\section{Inflationary Cosmology: various approaches} \label{sec:infl}

One of the big mathematical issues of a cosmology in general is an unpleasant nonlinearity of the cosmological equations involved. It is this very feature that makes the task of determining a dynamics of an inflationary universe a complicated and nontrivial endeavor, in many cases forcing the cosmologists to turn to various approximations. Even in the simplest case of a single scalar field $\phi$ with a potential $V(\phi)$, the dynamics of an early universe ends up being a system of two nonlinear equations:
\beq \label{FRW}
\begin{split}
&\ddot \phi  + 3 H \dot \phi + V'(\phi) = 0 \\
&H^2 = \frac{8 \pi}{3 M_p^2}\left(\frac{1}{2}\dot \phi^2 + V(\phi)\right) - \frac{k}{a^2},
\end{split}
\enq
where dot and prime indicate differentiation with respect to $t$ and $\phi$ correspondingly, $a$ is the scale factor and $H=\dot a / a$ is the Hubble parameter. There is a number of ways one might proceed with in order to simplify the system \eqref{FRW}. For example, one might introduce a new function $W(\phi)$, called a ``superpotential'' \cite{CZS}
\beqn
W=\frac{1}{2}\dot{\phi}^2+V(\phi),
\enqn
and in a special case of a flat universe with $k=0$, rewrite the system \eqref{FRW} as
\beqn
\begin{split}
\frac{d\phi}{dt}&=\mp\frac{\text{M}_{\text{P}}}{\sqrt{24\pi}}\frac{W'(\phi)}{\sqrt{W(\phi)}},\\
H & =\pm\frac{1}{\text{M}_{\text{P}}}\sqrt{\frac{8\pi}{3}W},
\end{split}
\enqn
enabling one to procure the solutions of \eqref{FRW} from a superpotential $W$ which is assumed to be known \cite{CZ,ZC}. Finally, in \cite{YY} it has been shown that the exact form of the superpotential can be derived from a known potential $V(\phi)$ via the formula
\beqn
W = V \left(1+\frac{1}{y^2-1}\right),
\enqn
with $y(\phi)$ being the solution of a famous Abel equation of the first kind:
\beq \label{Abel}
y'=-\frac{1}{2}\left(y^2-1\right)\left(\kappa - \chi' y\right),
\enq
where $\chi =\ln V$ and $\kappa = \pm 4 \sqrt{3\pi}/M_p$. Such a representation ends up being a versatile tool, particulary for analyzing the inflationary dynamics; for example, it is easy to see that for $V \ge 0$ the inflationary condition $\ddot a / a >0$ transforms into the condition $|y| > \sqrt{3}$ (cf. \cite{YYY}).

The problem with this method is, of course, the fact that even in the form \eqref{Abel} the cosmological equations remain, generally speaking, nonintegrable (a list of all known integrable cases in \cite{CR} has only 11 entries, and 7 of them contains no external parameters at all) -- which implies the non-integrability of the corresponding Friedmann equations. So, one can join an altogether different train of thought. Instead of trying to derive and study the exact solutions of either \eqref{FRW} or \eqref{Abel}, one might discard some of the smallish terms to get a new, simpler, completely integrable equation. This is the kind of idea that lies at the heart of a slow-rolling approximation. It is also the strategy that has been adopted in a recent article by James Lidsey \cite{L}. In it the author pursued an ambitious goal: to establish a new method of construction of general solutions for the universe that undergoes an inflationary expansion. The premise of the slow-rolling assumption is simple, for it assumes (correctly, as has been demonstrated in \cite{YYY}) that the for the major part of the inflationary phase the superpotential $W(\phi) \approx V(\phi)$, that is
\beq \label{slowrollcond}
2 | V(\phi) | /\dot \phi^2 \gg 1.
\enq
Rewriting the system \eqref{FRW} in a Hamilton-Jacobi form (where for the sake of brevity we utilize the rescaling $M_p=\sqrt{8 \pi}$)
\beq \label{HamJac}
\dot \phi = -2 H', \qquad V(\phi) = 3H^2 - 2 H'^2,
\enq
and leaving only the first two parameters from the Hubble slow roll hierarchy (for more information cf. \cite{K}) will, after some elementary calculations, yield a following second-order ODE:
\beq \label{slowroll}
4\frac{H''}{H}-8 \left(\frac{H'}{H}\right)^2=-\lambda^2,
\enq
with the parameter $\l^2 = 1 - n_s$, where $n_s < 1$ is called the slow-roll spectral index. Here Lidsey makes an interesting observation: a simple change of the variables
\beq \label{HF}
H^2(\phi) = 4 C F'(\phi)
\enq
where $C$ -- arbitrary constant, turns the left hand-side of \eqref{slowroll} into a Schwarzian operator
\beq \label{schwarz}
2\left(\frac{F''}{F'}\right)'-\left(\frac{F''}{F'}\right)^2 = -\lambda^2.
\enq
The benefit of this representation stems from the invariance of the Schwarzian differential operator with respect to a homographic transformation $f(\cdot) \to \frac{a f(\cdot) + b}{c f(\cdot) + d}$ with $a d - b c = 1$. Indeed, if $f(\phi)$ is the solution of \eqref{schwarz}, then due to the Schwarzian invariance, the function
\beq \label{inv}
F(\phi) = \frac{a f(\phi) + b}{c f(\phi) + d}, \qquad \left(\begin{array}{cc}
a & b \\
c & d
\end{array}\right) \in SL(2,\R),
\enq
would also solve \eqref{schwarz}. Now, since $F(\phi)=\eb^{\l \phi}$ serves as an obvious partial solution to \eqref{schwarz}, ``enhancing'' it by means of \eqref{inv} yields a whole new family of solutions of \eqref{schwarz}, and therefore, of \eqref{slowroll} (due to \eqref{HF}):
\beq \label{H2a}
H^2 = 4 \l C \frac{\eb^{\l \phi}}{(c \eb^{\l \phi} + d)^2}.
\enq
One can rewrite \eqref{H2a} in a physically sound manner by introducing a constant $H_0=H(0)$ and by subsequently getting rid of a constant $C$:
\beq \label{H2}
H^2 = H_0^2 \frac{(c + d)^2}{(c \eb^{\l \phi/2} + d \eb^{-\l \phi/2})^2}.
\enq
Thus, since $c$ and $d$ are two arbitrary constants, we evidently end up with a new method that seems to allow for a construction of a general solution of cosmological equations that describes the inflation under the slow-roll approximation. What's more, the method makes an extensive use of a Schwarzian derivative \eqref{schwarz}, whereas such a derivative is an integral tool in production of a travelling wave solution of Korteweg-de Vries equation via the B\"acklund transformation \cite{L}. So, it seems like we have indeed found a link, albeit an indirect one, between the cosmology and the KdV equation. Or did we instead unknowingly stumbled upon a remnant of something else?..

\section{The slow-roll condition and the Schr\"odinger equation} \label{sec:Sch}

Let us take one more look at the equation \eqref{slowroll}. First, let us rewrite it as:
\beq \label{H}
\left(\frac{H'}{H}\right)' - \left(\frac{H'}{H}\right)^2 = - \lambda^2/4.
\enq
An apparent transformation $A(\phi)=H'/H$ turns \eqref{H} into the Riccati equation:
\beq \label{Riccati}
A' - A^2 = - \lambda^2/4.
\enq
It is a well-known fact that a Riccati equation of the type \eqref{Riccati} is linearizable via the transformation $A(\phi) = - y'(\phi)/y(\phi)$, converting \eqref{Riccati} into the {\em Schr\"odinger equation}:
\beq \label{Sch}
y''(\phi) = \frac{\lambda^2}{4} y(\phi),
\enq
with the new variable $y$ related to a Hubble parameter $H$ as
\beqn
H = H_0\frac{y_0}{y}.
\enqn

Following \cite{L}, let us assume that the spectral parameter $\l$ is constant. The general solution of \eqref{Sch} turns out to be
\beqn
y(\phi) = C_1 \eb^{\l \phi/2} + C_2 \eb^{-\l \phi/2},
\enqn
and thus the Hubble parameter ends up being equal to
\beq \label{H3}
H(\phi)= H_0\frac{C_1 + C_2}{C_1 \eb^{\l \phi/2} + C_2 \eb^{-\l \phi/2}},
\enq
which, of course, is completely the same as the solution \eqref{H2}.

Now, let us stop for a second and contemplate the consequences of this result. One might be inclined to conclude, upon observing how undeniably identical the solutions \eqref{H2} and \eqref{H3} are, that the methods that produced those two solutions should therefore be treated as identical in all other meaningful respects. But we beg to differ. Firstly, we believe that working with the linear equation is more correct methodologically, especially since the alternative requires a nonlinear differential equation of not even second, but a {\em third} order \eqref{schwarz}. Secondly, and more importantly, the fact that the linearization delivers such a well-known and well-studied equation as \eqref{Sch} means that one might easily expand the scope of the model \eqref{slowroll} by allowing a slight time-dependence in the slow-roll spectral index $n_s$. Indeed, in the framework of the scalar-field problem the time-dependence can be replaced by a $\phi$-dependence. Thus, rescaling $\phi \to 2 x$ and subsequently rewriting the equation \eqref{slowroll} as
\beq \label{Sch2}
\frac{d^2 y(x)}{d x^2} = (1 - n_s(x)) y(x),
\enq
we end up with a standard Schr\"odinger equation with a potential $-n_s(x)$ and the spectral parameter $\mu = -1$. In order to illustrate this approach and see but two sample of the benefits it might offer, we will construct two particular integrable solutions that models the slow-rolling inflation such that the spectral index $n_s$ either stays close to $1$ for all sufficiently large $\mid \phi \mid$ or approaches the value $1$ in the process of inflation. We will use the Darboux transformation technique, which makes use of the fact that the equation \eqref{Sch2} is invariant with respect to the transformation $n_s \to \tilde n_s$ and $y \to \tilde y$ where
\beq \label{Darboux}
\tilde n_s = n_s + 2 (\ln y_1)'', \qquad \tilde y = y_2' - \frac{y_1'}{y_1} y_2,
\enq
and where $y_1$ and $y_2$ are two linearly independent solutions of Schr\"odinger equation \eqref{Sch2} (for further details see, for example, \cite{MS}, \cite{LY})\footnote{We would like to point out here that the Darboux transformation is not something entirely new for cosmologists. For example, it proved to be very useful in studies of the five-dimensional brane models -- see \cite{YY2}}.
\newline

{\bf Case 1. Dressing the Harrison-Zeldovich model.} Let us begin with the original, unperturbed equation
\beq \label{Dstep1}
\frac{d^2 y(x)}{d x^2} = y(x)- n_s y(x),
\enq
and let's assume that $n_s = 1$. This is a case of the Harrison-Zeldovich spectrum and it produces two linearly independent solutions of the form $y_1=a x + b$ and $y_2 = c x + d$ where $c b - a d = \Delta \neq 0$. Knowing them allows us to apply the Darboux transformation \eqref{Darboux}, receiving a {\em new} solution $\tilde y$ of the Schr\"odinger equation \eqref{Sch2}:
\beq \label{Dstep12}
\tilde y(x) = \frac{\Delta}{a x + b},
\enq
which corresponds to a {\em new} $\phi$-dependent spectral parameter:
\beq \label{Dstep13}
\tilde n_s(x) = 1 - 2\left(\frac{a}{ax+b}\right)^2.
\enq
From \eqref{Dstep12} it follows immediately that the Hubble parameter $H(\phi)= \frac{1}{2 \Delta} (a \phi + 2 b)$. The validity of the slow-roll condition \eqref{slowrollcond} for all large $\phi$ can then be checked directly from the Hamilton-Jacobi equations \eqref{HamJac}. Not surprisingly, it coincides with the region where $n_s \approx 1$. As $\phi$ approaches the value of $-2 b/a$, the potential of the scalar field $\phi$ gets closer to $\dot \phi^2/2$, eventually leading to a violation of the slow-roll condition -- and, therefore, of \eqref{H}, -- long before $n_s$ reaches infinity. We therefore conclude that the Darboux dressing of the Harrison-Zeldovich model leads to the inflationary model with a natural exit.
\newline

{\bf Case 2. Dressing the perturbed Harrison-Zeldovich model.} Let us now take a look at a situation where the original spectral parameter $n_s$ is close to but slightly smaller then $1$, i.e. when $\l^2 = 1-n_s > 0$. In this case the Schr\"odinger equation \eqref{Dstep1} has two linearly independent solutions $y_1=\cosh(\l x)$ and $y_2 = \sinh(\l x)$. After the Darboux transformation \eqref{Darboux} we will end up with a new, non-singular spectral parameter $\tilde n_s(\phi)$ defined as:
\beq \label{Dstep2}
\tilde n_s(\phi) = n_s +2 (1-n_s)~\sech^2(\sqrt{1-n_s} ~\phi /2).
\enq
Note, that the second term \eqref{Dstep2} is confined within the interval $(0, 2(1-n_s)]$, which can be chosen to be arbitrarily small by choosing an initial $n_s$ close to $1$, i.e. to the Harrison-Zeldovich spectrum. Let us illustrate this idea by applying it to the WMAP7+H0 data \cite{K2}. According to it, the observable spectral index lies in the interval $(0.939, 0.987)$ at the $2 \sigma$ confidence limit. Choosing $n_s=0.939$, we end up with $\tilde n_s(x) \in (0.939, 1.061)$. However, since the upper boundary is only reached in the limit $\phi \to \infty$, the real upper boundary must lie lower and correspond to some finite upper value of field $\phi$. The observational data provide such a value, for $n_s=0.987$ is reached when $\phi = 8.44$. Hence, the model not only predicts an inflation for large $\phi$ (which is checked directly as in Case 1), but also yields the exact value of the scalar field $\phi$ at which the inflation should have commenced as a direct consequence of observational data.

\section{Can KdV be relevant to cosmology?} \label{sec:Conclusion}

Before we conclude this article, we would like to return back to the question posed in the introductory section: does the KdV equation play any role in the inflationary dynamics of the universe? Or is it nothing but a distraction from an inherent structure of the equations involved? It seems that the arguments from section \ref{sec:Conclusion} turn the scales more to the favor of the latter. However, there is still one last aspect of the inflationary cosmology we have not yet discussed, and that is the possibility of a {\em multiple} scalar fields inflation - a concept that have received a fair share of attention lately (see, for example, \cite{LMS}, \cite{WBMR}, \cite{SL}). As before, let us assume that the primary mechanism for the inflation is the field $\phi$. However, this time we also assume that the universe is also filled by a new slowly-varying scalar field $\psi$, independent of $\phi$. Since by assumption it is {\em not} a primary agent of inflation, this means that a slow-rolling condition will still yield an equation similar to \eqref{H}, except that the slow-roll spectral index $n_s$ will now depend not just on $\phi$ but also on $\psi$. Conducting the linearization process as in the beginning of Section \ref{sec:Sch} will result in the following equation:
\beq \label{SchL}
\frac{d^2 y}{d x^2} = (1 - n_s(x, \psi)) y.
\enq
This is the one-parametric Schr\"odinger equation. If we want a spectrum of this problem (point $\{1\}$ in particular) to remain {\em independent} of field $\phi$ (and we do, if we want the right hand side of \eqref{SchL} to keep its original form), then we must impose a corresponding additional condition on $n_s(x, \psi)$. Such conditions are indeed known -- they form the hierarchy of differential equations called the {\em KdV hierarchy}, and, of course, the simplest of them is the Korteweg-de Vries equation itself \cite{L2}.

Thus, the final verdict on the question provided in the titles of this article is this: in the standard formalism of unique scalar field the inflation can be studied by means of a standard Schr\"odinger equation in the process that does not require any KdV machinery. However, in a situation of a multiple scalar fields either KdV or a equation from the KdV hierarchy indeed have a potential to become very relevant in our understanding of a spectral index of inflation.
\newline
\newline
{\bf Acknowledgement}

This work has been partially supported by project 14-02-31100 (RFBR, Russia)

\end{document}